\documentclass[twocolumn,aps, prb, final,superscriptaddress,showpacs]{revtex4}
\usepackage[latin9]{inputenc}
\usepackage{units}
\usepackage{amsmath}
\usepackage{amssymb}
\usepackage{esint}

\makeatletter

\providecommand{\tabularnewline}{\\}

\@ifundefined{textcolor}{}
{%
 \definecolor{BLACK}{gray}{0}
 \definecolor{WHITE}{gray}{1}
 \definecolor{RED}{rgb}{1,0,0}
 \definecolor{GREEN}{rgb}{0,1,0}
 \definecolor{BLUE}{rgb}{0,0,1}
 \definecolor{CYAN}{cmyk}{1,0,0,0}
 \definecolor{MAGENTA}{cmyk}{0,1,0,0}
 \definecolor{YELLOW}{cmyk}{0,0,1,0}
}

\usepackage{float}\usepackage{caption2}\usepackage{footnote}\usepackage{times}
\usepackage{graphicx}
\usepackage{times}

\makeatother

\begin{document}

\title{Terahertz dielectric response and coupled dynamics of ferroelectrics
and multiferroics from effective Hamiltonian simulations}

\author{Dawei Wang}

\address{Electronic Materials Research Laboratory, Key Laboratory of the Ministry
of Education and International Center for Dielectric Research, Xi'an
Jiaotong University, Xi'an 710049, China\\
 dawei.wang@mail.xjtu.edu.cn }

\author{Jeevaka Weerasinghe}

\address{Physics Department, University of North Texas, Denton, Texas 76205,
USA\\
 jeevaka.weerasinghe@unt.edu}

\author{Abdullah Albarakati }

\address{Physics Department, The University College at Al-Gammom, Umm Al-Qura
University Makkah, Saudi Arabia\\
 a\_albrak@yahoo.com}

\author{L. Bellaiche}

\address{Physics Department and Institute for Nanoscience and Engineering,
University of Arkansas, Fayetteville, Arkansas 72701, USA\\
 laurent@uark.edu}
\begin{abstract}
Ferroelectric and multiferroic materials form an important class of
functional materials. Over the last twenty years, first-principles-based
effective Hamiltonian approaches have been successfully developed
to simulate these materials. In recent years, effective Hamiltonian
approaches were further combined with molecular dynamics methods to
investigate \emph{terahertz dynamical} properties of various perovskites.
With this combination, a variety of ferroelectric and multiferroic
materials, including BaTiO$_{3}$, Ba(Sr,Ti)O$_{3}$, Pb(Zr,Ti)O$_{3}$,
BiFeO$_{3}$, and SrTiO$_{3}$ bulks and films have been simulated,
which led to the understanding of complex phenomena and discovery
of novel effects. In this review, we first provide technical details
about effective Hamiltonians and molecular dynamics simulation. Then,
we present applications of the combination of these two techniques
to different perovskites. Finally, we also briefly discuss possible
future directions of this approach. 
\end{abstract}


\maketitle

\section{Introduction \label{sec:Introduction}}

Ferroelectric materials have a spontaneous electric polarization that
can be reversed by applying an external electric field. These compounds
exhibit a wide range of important properties such as piezoelectricity,
high dielectric constants and photoelectricity, which make them key
materials for actuators, sensors, and potential candidates for energy
conversion and storage \cite{Lai2007,Prosandeev2007a,Ogihara2009}.
In recent years, multiferroic materials, which possess both spontaneous
electric polarization and magnetic orders also attracted much attention
\cite{Spaldin2010}. Moreover, the continuing miniaturization of electronic
devices has stimulated considerable research attention on ferroelectric
and multiferroic nanostructures, such as ultrathin films and nanodots,
which can have striking novel phenomena because of finite-size effects
\cite{Catalan2012}.

The investigation of ferroelectric materials can be approximately
divided into several stages. In 1920-1939, ferroelectricity were found
in Rochelle salt type crystals (KDP as an example) \cite{Cross1987}.
Then, after the discovery of BaTiO$_{3}$ in the early 1940s, the
number of studies on pure ferroelectric perovskites and solid solutions
grow rapidly \cite{Cross1987,Dove1997,Vijatovic2008}. In the 1960s
and 1970s, phenomenological Landau-type theories became widely used
to investigate phase transitions of perovskites, where the essential
idea is to express the free energy as Taylor expansions of order parameters
\cite{Lines1977,BruceA.D.andCowley1981,Salje}. At approximately the
same period, the soft mode theory for ferroelectrics was proposed
and turned out to be very useful \cite{Cochran1959,Cochran1960,Blinc1974,Scott1974,Cochran1981,BruceA.D.andCowley1981,Cross1987}.
Starting from the 1990s, the study of ferrelectric and multiferroic
materials was accelerated by huge increase of computational capacity
\cite{Phillpot2007} and subatomic scale observation techniques (e.g.,
high-resolution transmission electron microscopy) \cite{Jia2008,Nelson2011,Jia2011,Rossell2012}.

On the theoretical side, the development of first-principles-based
effective Hamiltonian approaches over the last 20 years proved to
be quite useful to mimic and understand finite-temperature properties
of perovskites (see Sec. \ref{sec:Effective-Hamiltonian-and}). More
recently, these approaches were combined with molecular dynamics to
predict complex dynamical properties of ferroelectrics and multiferroics,
in the terahertz frequency range (see Sec. \ref{sec:Applications-to-the}).

This review focuses on the combination of effective Hamiltonian approaches
and molecular dynamics methods, in general, and on the prediction
of properties resulting from its use, in particular. It is organized
as follows. In Sec. \ref{sec:Effective-Hamiltonian-and} and \ref{sec:Molecuar-dynamics-method},
we introduce the effective Hamiltonian approach and its implementation
within the molecular dynamics method. In Sec. \ref{sec:Applications-to-the},
we provide examples of the application of this approach to the investigation
of several perovskites. Finally, in Sec. \ref{sec:Conclusion}, a
summary is given and we invoke some possible future directions of
this approach.

\section{Effective Hamiltonian schemes, degrees of freedoms and energies \label{sec:Effective-Hamiltonian-and}}

Landau-type phenomenological theories have been used for more than
50 years to study ferroelectric materials, including their phase transitions
\cite{Chandra2007a}. On the other hand, the first microscopic, effective
Hamiltonian approach of ferroelectric materials was only developed
in the mid-nineties \cite{Zhong1994,King-Smith1994a,Zhong1995} although
the soft mode theory was known for a much longer time (see Refs. \cite{Cochran1959,Cowley1962}
and references therein). Since then, it has achieved great successes
in describing various ferroelectric materials \cite{King-Smith1994a,Zhong1995,Bellaiche2000,Bellaiche2002,Kornev2006,Kornev2007}.
In a nutshell, effective Hamiltonian methods first identify the most
important degrees of freedom in a ferroelectric material, and then,
based on symmetry arguments, construct the internal energy of the
system as a function of these degrees of freedom and their interactions.
The coefficients entering the effective Hamiltonian energy are typically
obtained by \textit{ab-initio} computations \cite{Zhong1994,Zhong1995}.
Initially, only the so-called local modes and strains of simple systems
were included in the effective Hamiltonian scheme \cite{King-Smith1994a,Zhong1995}.
However, later on, alloy effects were added. For instance, an effective
Hamiltonian scheme was developed to model (Ba,Sr)TiO$_{3}$ (BST)
solid solutions for the whole compositional range \textbf{\cite{Walizer2006}}.
Its total energy $E_{\textrm{tot}}$ has two main terms: 
\begin{eqnarray}
E_{\textrm{tot}} & = & E_{\textrm{VCA}}\left(\left\{ \mathbf{u}_{i}\right\} ,\left\{ v_{i}\right\} ,\left\{ \eta_{H}\right\} \right)\nonumber \\
 &  & +E_{\textrm{loc}}\left(\left\{ \mathbf{u}_{i}\right\} ,\left\{ v_{i}\right\} ,\left\{ \eta_{\textrm{loc}}\right\} ,\left\{ \sigma_{j}\right\} \right)\label{eq:energy-2}
\end{eqnarray}
where $\mathbf{u}\ensuremath{_{i}}$ denotes the local soft mode centered
on the Ti-site of the unit cell \emph{i} ($\mathbf{u}\ensuremath{_{i}}$
is directly proportional to the electric dipole of that cell); $\left\{ v_{i}\right\} $
are the dimensionless displacement variables of the cell corners and
are used to calculate inhomogeneous strain tensor components of the
cell \emph{i}; $\left\{ \eta_{H}\right\} $ is the homogeneous strain
tensor, which allows the simulation supercell to vary in size and
shape; $\sigma_{j}$ characterizes the atomic configuration, with
$\sigma_{j}$=+1 or -1 corresponding to the presence of a Ba or Sr
atom, respectively, at the A-lattice site \emph{j}; and $\left\{ \eta_{\textrm{loc}}\right\} $
represents the local strain resulting from the difference in ionic
size between Ba and Sr atoms, which is relatively large ($\simeq$2
\%). $E\mbox{\ensuremath{_{\textrm{VCA}}}}$ gathers the energy terms
solely involving the local soft mode, strain and their mutual couplings
resulting from the application of the virtual crystal approximation
\cite{Bellaiche2000,VCA2,Bellaiche2000a} to model (Ba$_{0.5}$Sr$_{0.5}$)TiO$_{3}$
solid solutions. On the other hand, $E_{\textrm{loc}}$ can be thought
of as a perturbative term due to the fact that BST systems possess
real Ba and Sr atoms on the A-sites rather than a virtual, compositional-dependent
$\left\langle A\right\rangle $ atom.

\begin{figure}[h]
\noindent \begin{centering}
\includegraphics[width=3.5cm]{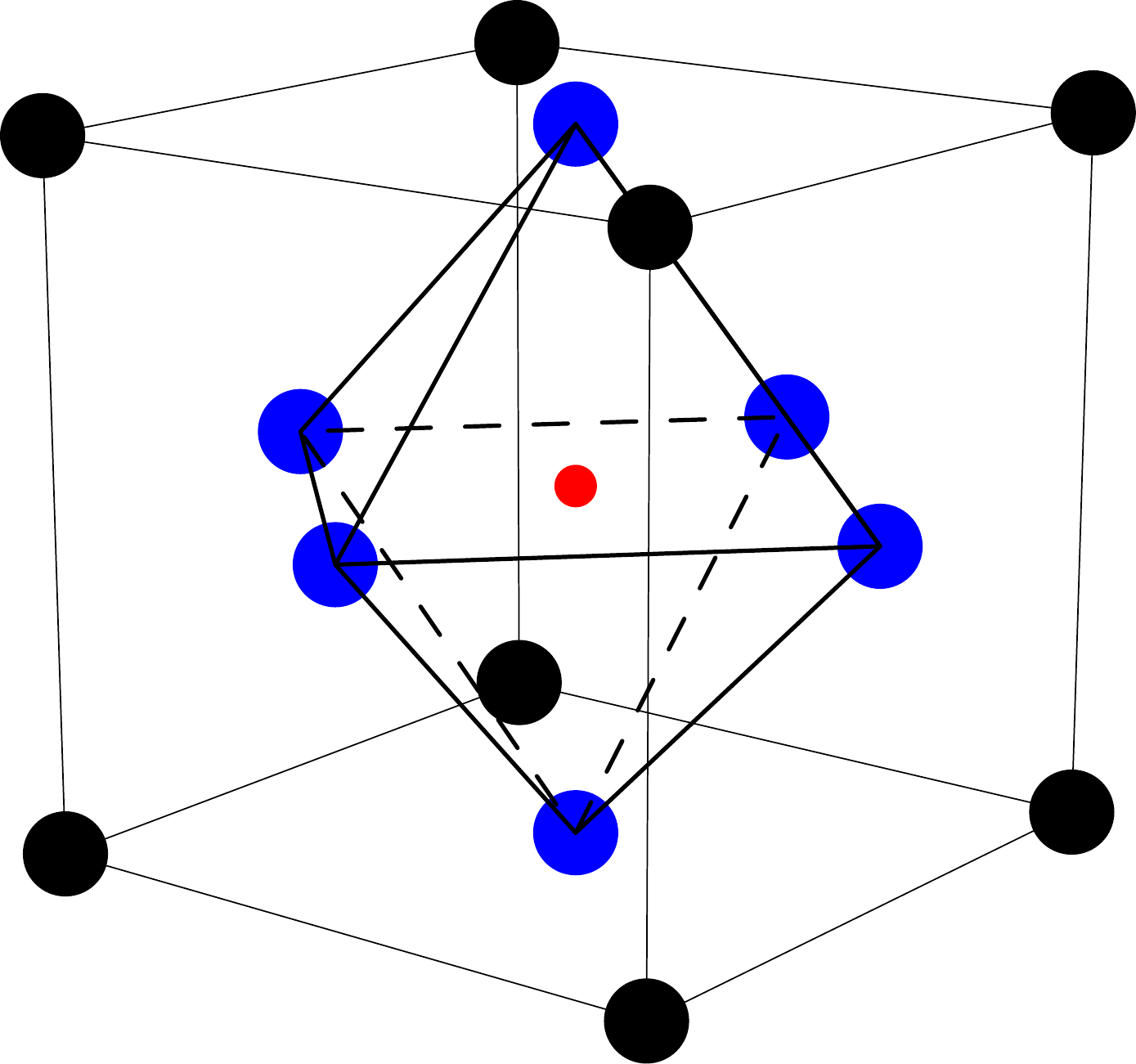}\includegraphics[width=3.5cm]{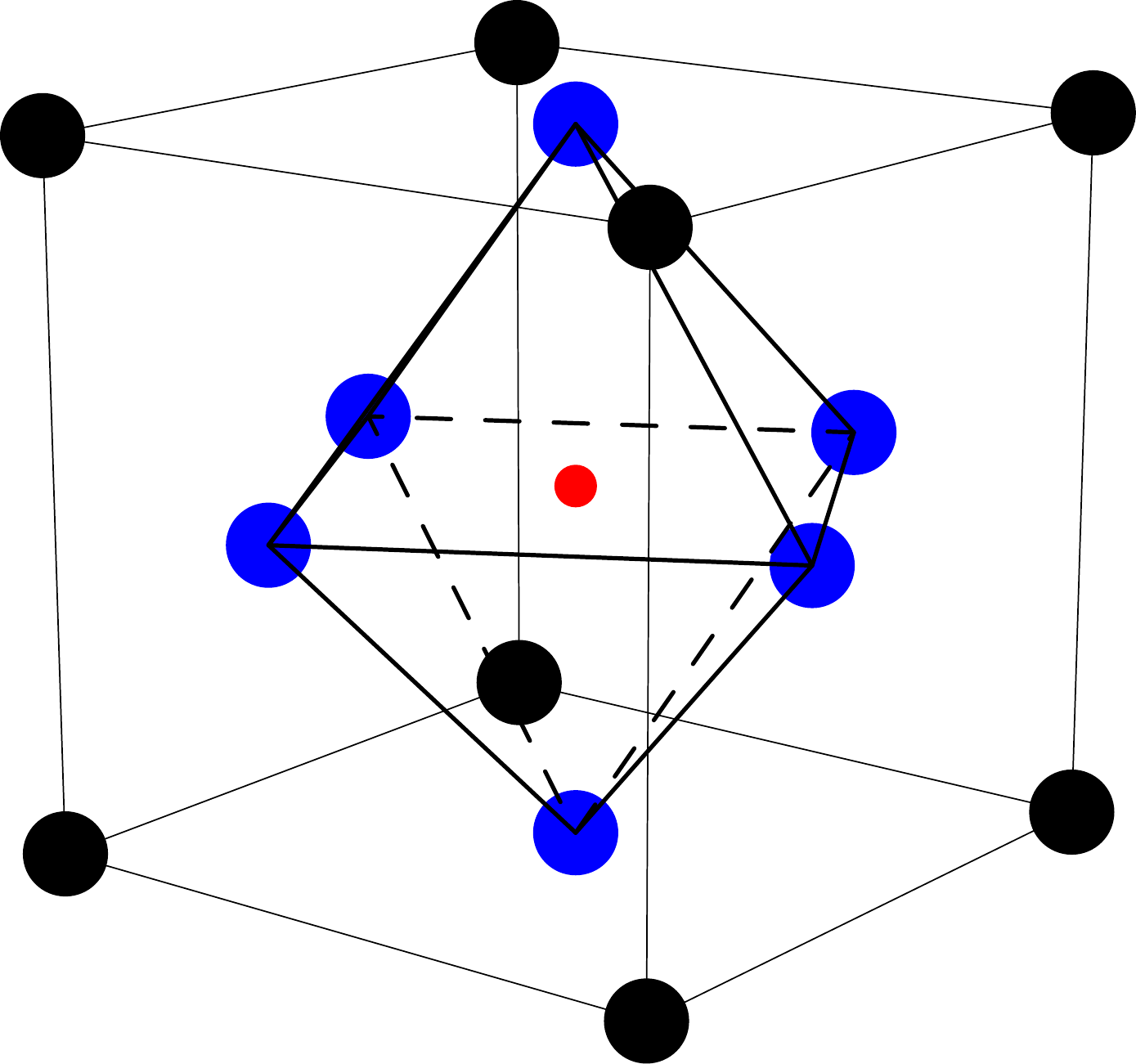} 

\par\end{centering}

\caption{For an ABO$_{3}$ system, the black dots (on the corners), red dots
(in the center) and the blue dots (on the face centers) represent
the A, B and oxygen atoms, respectively. The left panel shows the
atoms at the ideal positions of the cubic phase, while the right panel
shows the oxygen octahedron rotated around the pseudo-cubic $\left[111\right]$.
\label{fig:AFD-Rot}}
\end{figure}

Antiferrodistortive (AFD) motions \cite{Vanderbilt1998,Kornev2006}
and magnetic degrees of freedom \cite{Kornev2007,Albrecht2010} were
also incorporated in some effective Hamiltonians. For instance, in
the multiferroic BiFeO$_{3}$ material, the resulting effective Hamiltonian
has a total energy that can be written as a sum of two main terms
\cite{Kornev2007,Albrecht2010,Rahmedov2012}:

\begin{align}
E_{\textrm{tot}}= & E_{\textrm{FE-AFD}}(\{{\bf \mathbf{u}_{{\it i}}}\},\{\eta\},\{{\bf \mathbf{\omega}_{{\it i}}}\})\nonumber \\
 & +E_{\textrm{MAG}}(\{{\bf \mathbf{m}_{{\it i}}}\},\{{\bf \mathbf{u}_{{\it i}}}\},\{\eta\},\{{\bf \mathbf{\omega}_{{\it i}}}\}),\label{eq:Etot}
\end{align}
where the ${\bf \mathbf{\omega}_{{\it i}}}$ pseudo-vector characterizes
the tilting of oxygen octahedra in unit cell $i$ of the investigated
perovskite \cite{Kornev2007} (see Fig. \ref{fig:AFD-Rot}). Note
that such tilting is also termed the antiferrodistortive (AFD) motion.
$\mathbf{m}_{i}$ is the magnetic dipole moment centered on the Fe-site
$i$ and has a fixed magnitude of 4$\mu_{B}$ \cite{Neaton2005}.
$E_{\textrm{FE-AFD}}$ is given in Ref. \cite{Kornev2006} and involves
terms associated with ferroelectricity, strain and AFD motions, and
their mutual couplings. $E_{\textrm{MAG}}$ gathers magnetic degrees
of freedom and their interactions with local modes, strains and oxygen
octahedral tiltings, and is given in Refs. \cite{Kornev2007,Albrecht2010,Rahmedov2012}.

\section{Combining Molecular dynamics and Effective Hamiltonians\label{sec:Molecuar-dynamics-method}}

Effective Hamiltonians can be put into Monte-Carlo (MC) algorithms
(see, e.g. Refs.\cite{Zhong1994,Bellaiche2000,Kornev2007}) or molecular
dynamics (MD) simulations (see, e.g., Refs. \cite{Ponomareva2008a,Wang2011}),
to predict material properties. A MC simulation generates configurations
of a system with certain probabilities determined by their free energies
at certain temperatures. The generated configurations are then used
to find physical properties of the system statistically. The MC method
has been put in use for finding static properties of ferroelectric
materials since the introduction of the first effective Hamiltonian
method \cite{Zhong1994}. Examples include its application to BaTiO$_{3}$
\cite{Zhong1994}, Pb(Zr,Ti)O$_{3}$ solid solutions \cite{Bellaiche2000,Bellaiche2002,Kornev2006}
and BiFeO$_{3}$ mulltiferroics \cite{Kornev2007,Albrecht2010,Rahmedov2012,Prosandeev2012a},
for which the MC algorithm accurately predicted some important physical
properties at finite temperature.

On the other hand, the effective-Hamiltonian-based MD method deals
with the \textit{dynamics} of the degrees of freedom directly, employing
Newton's equation for the local modes, AFD motions and strains for
ferroelectrics \cite{Ponomareva2008a,Wang2011}, and also the Landau-Lifshitz-Gilbert
equation for magnetic moments \cite{Garcia-Palacios1998,Gilbert2004,Wang2012}
in multiferroics. This approach was used as early as the end of the
1990s to investigate KNbO$_{3}$ perovskites \cite{Krakauer1999}.
Later, the method was further developed by Ponomareva \emph{et al}
\cite{Ponomareva2008a} to obtain terahertz dielectric response of
BaTiO$_{3}$ (which is a simple system where the two degrees of freedom
of the effective Hamiltonian are local modes and strains). Surprisingly,
two peaks (instead of only one single peak) were numerically found
in the THz dielectric response \cite{Ponomareva2008a}. At this point,
MD already shows its usefulness in understanding subtle phenomena.
The effective-Hamiltonian-based MD method was further developed to
include the dynamic variable related to AFD motions to investigate
Pb(Zr,Ti)O$_{3}$ \cite{Wang2011,Wang2011b,Weerasinghe2012}. Recently
the magnetic degree of freedom was also added into the combination
of MD technique and effective Hamiltonians to investigate terahertz
dynamical properties of multiferroics \cite{Wang2012}.

Combining MD and effective Hamiltonians has allowed the computation
of complex dynamical properties, including dielectric responses \cite{Ponomareva2008a,Wang2011,Wang2011b,Wang2012},
electrocaloric effect \cite{Lisenkov2009a,Beckman2012,Ponomareva2012},
domain dynamics \cite{Zhang2011b}, the creation of electric field
by dynamically altering the magnetic dipolar configurations in ferromagnets
\cite{Prosandeev2009}, and the investigation of BaTiO$_{3}$ thin-film
capacitors \cite{Nishimatsu2008}. Let us now describe in details
this combination.

\subsection{Dynamical equations obeyed by the structural degrees of freedom}

The \emph{structural} degrees of freedom of the effective Hamiltonian
obey Newton's equation, that is:\cite{Ponomareva2008a,Wang2011}

\begin{align}
M_{A_{\alpha}}\frac{d^{2}}{dt^{2}}A_{\alpha}= & -\frac{\partial}{\partial A_{\alpha}}E_{\textrm{tot}},\label{eq:newton-equation-1}
\end{align}
where $A_{\alpha}$ is one (Voigt or Cartesian) component of the aforementioned
local modes, strains, and/or AFD vectors. $E_{\textrm{tot}}$ is the
internal energy provided by the effective Hamiltonian scheme for the
system under consideration (e.g. Eq. (\ref{eq:Etot}) for BiFeO$_{3}$).
$M_{A_{\alpha}}$ has a dimension of a mass for the local modes and
strains, while it can be interpreted as a moment of inertia when $A_{\alpha}$
is a Cartesian component of the AFD vectors. Note that, during the
MD simulations \cite{Ponomareva2008a,Wang2011}, the temperatures
of all these structural degrees of freedom are controlled by Evans-Hoover
thermostats \cite{Evans1983}.

\subsection{Treating magnetic moments on a dynamical point of view \label{sub:Treating-magnetic-moments}}

In multiferroic materials, such as BiFeO$_{3}$, there exists yet
another important degree of freedom: the magnetic moment. In contrast
to the structural degrees of freedom, the dynamics of magnetic moments
is described by the Landau-Lifshitz-Gilbert (LLG) equation \cite{Gilbert2004}.
The implementation and incorporation of LLG into MD simulations for
multiferroics is not trivial since these simulations should also include
the Newtonian dynamics followed by the structural degrees of freedom.
This explains why many studies treat the magnetic moments and the
structural degrees of freedom separately \cite{Garcia-Palacios1998,Ma2008,Ma2010},
showing the challenge of merging these two dynamics (even more so
at finite temperature).

This merging was, in fact, accomplished in Ref. \cite{Wang2012} for
multiferroic BiFeO$_{3}$ using the effective Hamiltonian approach.
For that, the stochastic Landau-Lifshitz-Gilbert (LLG) equation was
implemented for the $\mathbf{m}_{i}$, the magnetic moment on unit
cell $i$ \cite{Garcia-Palacios1998}: 
\begin{align}
\frac{d\mathbf{m}_{i}}{dt}= & -\gamma\mathbf{m}_{i}\times\left[\mathbf{B}_{\textrm{eff}}^{i}\left(t\right)+\mathbf{b}_{\textrm{fl}}^{i}\left(t\right)\right]\nonumber \\
 & -\gamma\frac{\lambda}{\left|\mathbf{m}_{i}\right|\left(1+\lambda^{2}\right)}\mathbf{m}_{i}\times\left\{ \mathbf{m}_{i}\times\left[\mathbf{B}_{\textrm{eff}}^{i}\left(t\right)+\mathbf{b}_{\textrm{fl}}^{i}\left(t\right)\right]\right\} ,\label{eq:stochastic-LLG}
\end{align}
where $\mathbf{B}_{\textrm{eff}}^{i}=-\nabla_{\mathbf{m}_{i}}E_{\textrm{tot}}$
is the effective magnetic field acting on the $i$th magnetic moment,
$\gamma$ is the gyromagnetic ratio, $\lambda$ is the damping coefficient.
$\mathbf{b}_{\textrm{fl}}^{i}$ is a fluctuation field that also acts
on the $i$th magnetic moment, and allows to control the temperature
of magnetic moments, as consistent with previous studies done on magnetic
systems \cite{Brown1963,Kubo1970}. The Box-Muller method was used
in Ref. \cite{Wang2012} to generate random numbers obeying Gaussian
distribution for $\mathbf{b}_{\textrm{fl}}^{i}$ and to enforce the
following conditions obeyed by this fluctuation field at the finite
temperature, $T$ \cite{Brown1963,Garcia-Palacios1998}: 
\begin{align}
\left\langle \mathbf{b}_{\textrm{fl}}^{i}\right\rangle = & 0,\label{eq:bfl-1st-moment-1}\\
\left\langle b_{\textrm{fl},\alpha}^{i}\left(t_{1}\right)b_{\textrm{fl},\beta}^{i}\left(t_{2}\right)\right\rangle = & 2\frac{\lambda k_{B}T}{\gamma\left|\mathbf{m}_{i}\right|}\delta_{\alpha,\beta}\delta\left(t_{1}-t_{2}\right),\label{eq:bfl-2nd-moment-1}
\end{align}
where $\alpha$ and $\beta$ denote Cartesian coordinates and $t_{1}$
and $t_{2}$ are two different times. $\left\langle \right\rangle $
indicates an average over possible realizations of the fluctuating
field \cite{Garcia-Palacios1998},\textbf{ }$\delta_{\alpha,\beta}$
is the Kronecker delta function and $\delta\left(t_{1}-t_{2}\right)$
is the Dirac delta function.

Note that (i) the LLG equation shown in Eq. (\ref{eq:stochastic-LLG}),
which contains the fluctuation field, is a Stratonovich stochastic
differential equation \cite{Garcia-Palacios1998}; and (ii) the semi-implicit
method devised by Mentink \emph{et al }\cite{Mentink2010} was adopted
in Ref. \cite{Wang2012} to properly integrate this LLG equation.
This algorithm was found to be efficient and accurate enough for our
investigations of multiferroics.

\subsection{Obtaining dynamical responses via correlation function}

For ferroelectric and multiferroic materials, the complex dielectric
response, $\varepsilon_{\alpha\beta}\left(\nu\right)$, can be obtained
via the proposed MD scheme in the gigahertz-terahertz regime at any
temperature, using the following equation \cite{Caillol1986,Ponomareva2008a,Hlinka2008}:
\begin{align}
\varepsilon_{\alpha\beta}\left(\nu\right)-1= & \frac{1}{\varepsilon_{0}Vk_{B}T}\left[\left\langle d_{\alpha}(t)d_{\beta}(t)\right\rangle +\right.\nonumber \\
 & \left.i2\pi\nu\int_{0}^{\infty}dte^{i2\pi\nu t}\left\langle d_{\alpha}\left(t\right)d_{\beta}\left(0\right)\right\rangle \right]\label{eq:dielectric-response-from-MD}
\end{align}
where $\nu$ is the frequency while $\alpha$ and $\beta$ define
Cartesian components and $V$ is the volume of the chosen supercell,
$\mathbf{d}\left(t\right)$ is the dipole moment of the system at
time $t$, and $\left\langle ...\right\rangle $ represents thermal
averages.

Typically, each peak found in simulations in the spectra of $\varepsilon_{\alpha\beta}\left(\nu\right)$
is fitted by a classical harmonic oscillator 
\begin{align}
\varepsilon_{\alpha\beta}\left(\nu\right)= & \frac{S\nu_{r}^{2}}{\nu_{r}^{2}-\nu^{2}+i\nu\gamma},\label{eq:DHO}
\end{align}
where $\nu_{r}$, $\gamma$ and $S$ are the resonant frequency, damping
constant, and strength of the oscillator, respectively. In some cases
(e.g., when a central mode exists), other spectra models have also
been used \cite{Weerasinghe2013}.

Other type of dynamical responses can also be obtained. For example,
one can replace the dipole moment by the magnetization or an AFD vector
in Eq.(\ref{eq:dielectric-response-from-MD}) to predict the complex
magnetic susceptibility \cite{Wang2012} or the physical response
of AFD vectors to their $ac$ conjugate field, respectively.

\section{Applications to the study of a variety of systems \label{sec:Applications-to-the}}

Having discussed the technical side of the effective-Hamiltonian MD
approach for complex dynamical properties, we now show its application
to ferroelectric and multiferroic materials via several examples.

\subsection{Soft mode and central mode in BaTiO$_{3}$ and Ba(Sr,Ti)O$_{3}$
systems \label{sub:Spectra-of-BaTiO}}

BaTiO$_{3}$ is the first ferroelectric material that was studied
in detail using the effective Hamiltonian method \cite{Zhong1994,Zhong1995}.
It is therefore not surprising that the MD method was also first developed
to obtain the complex dielectric response of BaTiO$_{3}$, which has
``only'' local modes and strains in the effective Hamiltonian. This
``simple'' system, however, gave rise to a surprise: it was numerically
found that, for any temperature ranging between the Curie temperature
$T_{C}$ and 750 K (the system is in the paraelectric phase), a good
fit of the dielectric response requires \textit{two} classical damped
harmonic oscillators rather than a single one \cite{Ponomareva2008a}.
These two different overdamped modes of comparable strengths were
found to contribute to the total dielectric response. The lower frequency
mode softens as temperature approaches the Curie point from above.
The higher-frequency mode softens less and stays in the 60\textendash{}100\,$\textrm{cm}^{-1}$
range and was found to be associated with temperature-independent
short-range correlations.

Similar results were also found in (Ba$_{x}$Sr$_{1-x}$)TiO$_{3}$
with $x>0.4$, but not in Pb(Zr$_{0.4}$,Ti$_{0.6}$)O$_{3}$ \cite{Ponomareva2008a}.
It was also found in the tetragonal ferroelectric phase of BaTiO$_{3}$
\cite{Hlinka2008}. There, it was indicated that the low-frequency
mode (which is often referred to as the ``central mode'' , CM) is
related to the jumping of the local mode between different, but equivalent,
potential minimum. This differs from the high-frequency mode, which
is the well-known soft mode (SM) and that corresponds to dipolar oscillations
around one energy minimum. Note that the CM in tetragonal BaTiO$_{3}$
was analyzed using a Debye relaxation model, and that the characteristic
frequency of this mode was found to decrease as the temperature approaches
$T_{C}$ from below \cite{Hlinka2008}.

More recently, MD simulations also helped to elucidate the dielectric
response in the paraelectric phase of (Ba, Sr)TiO$_{3}$ solid solutions,
and to find whether the CM persists up to the highest temperatures,
or rather if it disappears at some well-defined temperature $T_{\textrm{CM}}$
that is higher than the Curie temperature \cite{Ostapchuk2009,Weerasinghe2013a}.
To account simultaneously for the SM and CM appearing in the dielectric
response, the following equation was used \cite{Hlinka2008,Kadlec2009a}:
\begin{eqnarray}
\varepsilon\left(\nu\right) & = & \frac{S}{\nu_{0}^{2}-\nu^{2}-i\nu\Gamma-\nicefrac{\delta^{2}}{(1-\nicefrac{i\nu}{\nu_{R}})}}\label{eq:5}
\end{eqnarray}
where $S$, $\nu_{0}$, $\Gamma$ are the oscillator strength, frequency
and damping constant of the SM, respectively; $\nu_{R}$ and $\delta$
are the relaxation frequency of the CM, and the coupling coefficient
between the SM and CM, respectively. The MD results suggested that
the coupling term $\delta$ is constant, while $\nu_{R}$ increases
with temperature. Comparing three different models (that differ about
the assumptions on the dependence of $\delta^{2}$ and $\nu_{R}$
on temperature) to MD simulation results indicated a thermally activated
dynamics of the coupling coefficient of CM-SM. It was shown that Eq.
(\ref{eq:5}) can be used for all temperatures with well obtained
parameters, and that the appearance of the CM in the dielectric spectra
can be understood as a crossover between two regimes: a higher-temperature
regime governed by the soft mode only (when $\nu_{R}\gg\delta^{2}/\Gamma$)
\textit{versus} a lower-temperature regime exhibiting a coupled soft
mode/central mode dynamics. The crossover temperature $T_{\textrm{CM}}$
can be quantitatively determined from the fitting parameters of Eq.
(\ref{eq:5}).

Moreover, the central mode had also been observed in SrTiO$_{3}$
films deposited on DyScO$_{3}$ substrates under tensile strain \cite{Kadlec2009a,Nuzhnyy2009}
and was shown to have a strong effect on tunability \cite{Weerasinghe2013}
(see Sec. \ref{sub:Tunability-of-SrTiO}).

\subsection{Bilinear and nonlinear dynamical couplings in Pb(Zr,Ti)O$_{3}$ \label{sub:Linear-and-nonlinear}}

\subsubsection{Bilinear coupling \label{sub:Linear-coupling}}

In contrast to BaTiO$_{3}$, Pb(Zr,Ti)O$_{3}$ possesses antiferrodistortive
oxygen octahedra tiltings that can strongly affect its properties.
For example, it was shown that the correct Curie temperature can be
obtained only when AFD is included as an independent degree of freedom
in the effective Hamiltonian \cite{Bellaiche2000a,Kornev2006}. Moreover,
it was found that, below some critical temperature ($\simeq$ 200\,K,
the exact value depends on the Ti composition), the AFD tiltings adopt
a long-range order, which is described by the following two equations:

{\small{
\begin{align}
\left\langle \boldsymbol{\omega}\right\rangle = & \frac{1}{N}\sum_{i}\boldsymbol{\omega}_{i}=0,\label{eq:average}\\
\left\langle \boldsymbol{\omega}_{R}\right\rangle = & \frac{1}{N}\sum_{i}\boldsymbol{\omega}_{i}\left(-1\right)^{n_{x}\left(i\right)+n_{y}\left(i\right)+n_{z}\left(i\right)}\neq0.\label{eq:R-average}
\end{align}
}}where $\boldsymbol{\omega}_{i}$ is the tilting angle of the oxygen
octaheron at site $i$. These two equations show that, while the simple
average of the tilting angle over the whole system is zero, the average
associated with the $R$-point of the cubic Brillouin (i.e., multiplying
by $-1$ when going from one B-site to any of its first nearest neighbor
B sites) is finite.

Since an additional degree of freedom is identified in Pb(Zr,Ti)O$_{3}$,
its dynamical coupling to the electric polarization is of interest.
For instance, one can wonder if this coupling can give rise to additional
peaks in Raman spectra or to other unexpected phenomena. An effective-Hamiltonian-based
MD approach was further developed and exploited to address this important
topic \cite{Wang2011,Wang2011b}. First, the energy terms related
to the AFD degree of freedom were incorporated in the MD program.
Among others, the energy term representing the interaction energy
between the local mode and AFD tiltings is given by 
\begin{align}
E_{\textrm{AFD-FE}}= & \sum_{i}\sum_{\alpha,\beta,\gamma,\delta}D_{\alpha,\beta,\gamma,\delta}\omega_{i,\alpha}\omega_{i,\beta}u_{i,\gamma}u_{i,\delta},\label{eq:FE-AFD-couplings}
\end{align}
where $i$ runs over all the unit cells and $\alpha,\beta,\gamma$
and $\delta$ denote Cartesian components, with the $x$, $y$, and
$z$ axes being chosen along the pseudocubic {[}100{]}, {[}010{]},
and {[}001{]} directions, respectively. The $D_{\alpha,\beta,\gamma,\delta}$
matrix elements quantify the couplings between local modes and AFD
motions, and are extracted from local-density-approximation (LDA)
calculations \cite{Hohenberg1964,Bellaiche2000}. Second, in addition
to the dielectric response of Eq. (\ref{eq:dielectric-response-from-MD}),
the MD approach was also able to compute the following quantity: 
\begin{align}
\varepsilon_{\alpha\beta}^{\textrm{AFD}}\left(\nu\right)-1= & \frac{1}{\varepsilon_{0}Vk_{B}T}\left[\left\langle \omega_{R,\alpha}(t)\omega_{R,\beta}(t)\right\rangle +\right.\nonumber \\
 & \left.i2\pi\nu\int_{0}^{\infty}dte^{i2\pi\nu t}\left\langle \omega_{R,\alpha}\left(t\right)\omega_{R,\beta}\left(0\right)\right\rangle \right],\label{eq:response-expression-AFD}
\end{align}
where $\boldsymbol{\omega}_{R}\left(t\right)$ are the order parameter
vector associated with the AFD motions at the $R$-point of the cubic
Brillouin zone \cite{Kornev2006} {[}see Eq. (\ref{eq:R-average}){]}
at time $t$ and $\varepsilon_{\alpha\beta}^{\textrm{AFD}}\left(\nu\right)$
can be thought as the response of $\mathbf{\omega_{R}}\left(t\right)$
to its $ac$ conjugate field, which is a time-dependent staggered
field. This physical response is then fitted by classical harmonic
oscillators, which allows the identification of the natural resonant
frequencies of the AFD motions.

Applying this newly developed MD approach to Pb(Zr$_{0.55}$Ti$_{0.45}$)O$_{3}$,
a cubic paraelectric phase was found for temperatures above the Curie
temperature $T_{C}\sim700$\,K, and a $R3m$ phase, in which the
polarization is along a $\left\langle 111\right\rangle $ direction,
exists between $T_{C}$ and another critical temperature $\sim200$\,K.
Below $\sim200$\,K, the oxygen octahedra tilting has a long-range
ordering, rotating around the same $\left\langle 111\right\rangle $
direction of the polarization, setting the system in the $R3c$ phase.
These results are in good agreement with experiments \cite{Jaffe1971,Noheda2000,Noheda2002}
and previous MC simulations \cite{Kornev2006}. Then, the dielectric
response of the system was obtained, and the resonant frequencies
of the lowest-in-frequency dielectric peak {[}an $E$(1TO) mode{]},
$\nu_{r}$, were extracted. Such a mode represents dipolar oscillations
perpendicular to the spontaneous polarization. Its frequency versus
temperature follows the relation $\nu_{r}=C\left|T-T_{C}\right|^{1/2}$
close to $T_{C}$, which can be explained by Landau theory and which
is an expected behavior. However, it was also found that two peaks,
rather than a single $E$(1TO) peak, exist in the 50-70\,$\textrm{cm}^{-1}$
range for temperatures below $\sim200$\,K, which is the transition
temperature at which the $R3m$ to $R3c$ phase transition occurs,
which agrees well with Raman experimental findings \cite{Bauerle1976,Deluca2010}. 

In order to understand the origin of this additional mode, (i) the
coupling between the local mode and AFD was switched off by setting
$D_{\alpha,\beta,\gamma,\delta}$, to zero, resulting in only one
dielectric peak remaining below $\sim200$\,K ; (ii) the natural
frequencies of oxygen octahedra tiltings were extracted from the analysis
of Eq. (\ref{eq:response-expression-AFD}), leading to the discovery
that the AFD degree of freedom clearly has two natural frequencies
in the 50\textendash{}75\,$\textrm{cm}^{-1}$ range below 200\,K
that coincides with the two peaks observed in the dielectric response.
These findings point to the fact that the coupling between the AFD
and local mode is the reason for the additional peak below 200\,K.
Due to this coupling, the AFD acquires some polarity, while the ``usual\textquotedblright{}
$E$ dielectric mode loses some polar character, resulting in the
decrease of the electric dipole spectral weight below 200 K. When
the system is in the $R3c$ phase or other similar phases ($\left\langle \mathbf{u}\right\rangle \neq0$
and $\left\langle \boldsymbol{\omega_{R}}\right\rangle \neq0$, see
below), the coupling between local mode and AFD motions can be expressed
as 
\begin{align}
H_{\textrm{coupl,dynam}}\simeq & \sum_{i}\kappa\left|\left\langle \mathbf{u}\right\rangle \right|\left|\left\langle \boldsymbol{\omega}_{R}\right\rangle \right|\tilde{\mathbf{u}}_{i}\tilde{\boldsymbol{\omega}}_{i},\label{eq:linear-coupling}
\end{align}
where $\left|\dots\right|$ denotes the magnitude of a vector, $\tilde{\mathbf{u}}_{i}$
(respectively, $\tilde{\boldsymbol{\omega}}_{i}$) is the deviation
of the local mode (respectively, AFD mode) in the unit cell $i$ with
respect to their spontaneous values. It was verified that the behavior
of the spectra below 200\,K can be understood by analyzing the dynamical
equation for $\tilde{\mathbf{u}}_{i}$ under the influence of AFD,
which is derived from the coupling of Eq. (\ref{eq:linear-coupling}).

The effect of composition on the low-frequency modes at low temperature
was also investigated \cite{Weerasinghe2012}. At this point it is
important to know that, within a small range of compositional variation
(45\%-56\% Ti), the morphotropic boundary of Pb(Zr,Ti)O$_{3}$ supports
three low-temperature phases, namely rhombohedral $R3c$, monoclinic
$Cc$, and tetragonal $I4cm$. This results in the rotation of the
polarization from the {[}111{]} to {[}001{]} pseudo-cubic direction
as the composition varies, which is the underlying reason why Pb(Zr,Ti)O$_{3}$
has good performance in this region \cite{Jaffe1971}. Via the application
of MD simulations, the number of low-frequency modes $\left(<89\,\textrm{cm}^{-1}\right)$,
their resonant frequencies, and dielectric spectral weights were obtained
to predict some characteristics of the low-frequency optical modes
that can appear in the dielectric and Raman spectra at low temperature.
Throughout this compositional range, both the local mode and AFD have
long-range orders at low temperature, therefore the coupling of local
mode and AFD is an important issue in analyzing the simulation results.

Practically, MD simulations were carried out for disordered PZT solid
solutions, with Ti compositions ranging from 45.2\% to 56.0\% in intervals
of 0.2\% and for temperatures ranging from 1100K down to 10 K. The
results are summarized in Table 1.

\begin{widetext}\savenotes 
\begin{table}[h]
\noindent \begin{centering}
\begin{tabular}{cccc}
\hline 
Ti Composition  & 45.2\%--47.5\%  & 47.5\%--51.0\%  & 51.0\%--56.0\%\tabularnewline
\hline 
\hline 
Phase  & rhombohedral $R3c$  & monoclinic $Cc$  & tetragonal $I4cm$\tabularnewline
\hline 
Number of modes  & 2%
\footnote{Two (double degenerate) $E$ modes.%
}  & 4 %
\footnote{The degeneracy between the $E$ modes is lifted.%
}  & 2 %
\footnote{Two (double degenerate) $E$ modes.%
} \tabularnewline
\hline 
Origin  & local mode and AFD \cite{Wang2011}  & local mode and AFD  & local mode and AFD\tabularnewline
\hline 
\end{tabular}
\par\end{centering}

\caption{Phases and low-frequency modes through the morphotropic boundary of
Pb(Zr,Ti)O$_{3}$ at 10K.\textbf{ }}
\end{table}

\spewnotes\end{widetext}

As indicated in Table 1, different phases have different number of
modes. The $R3c$ phase has two modes ($E^{\left(1\right)}$ and $E^{\left(2\right)}$),
the $Cc$ phase has four modes ($A^{\prime\left(1\right)},\, A^{\prime\prime\left(1\right)},\, A^{\prime\left(2\right)}$
and $A^{\prime\prime\left(2\right)}$), and the $I4cm$ again has
two modes ($E^{\left(1\right)}$ and $E^{\left(2\right)}$), which
are consistent with group theory. It was found that the frequency
differences between the $A^{\prime}$ and $A^{\prime\prime}$ modes
originating from $E^{\left(1\right)}$ and $E^{\left(2\right)}$,
are largest near the compositional midpoint of the $Cc$ phase (that
is located around a Ti composition of 49.4\%) and decrease to either
side of it. That is, the splitting of the $E$ mode in the monoclinic
phase depends on how strong the phase is monoclinic, which can be
quantified by the monoclinic depth parameter defined in Eq. (4) of
Ref. \cite{Weerasinghe2012}. Analytical model was constructed, based
on the coupling of Eq. (\ref{eq:FE-AFD-couplings}), to fit the frequencies
of each mode and understand the variation of spectra weight with respect
to Ti composition at different phases. In particular, the splitting
in the $Cc$ phase can be fitted with monoclinic depth as a fitting
parameter. Such analytical models were able to accurately reproduce
characteristics of the low-frequency optical modes and led to a better
understanding of the coupling between ferroelectric and AFD degrees
of freedom.

\subsubsection{Nonlinear coupling}

The coupling between the local mode and AFD as shown in Eq. (\ref{eq:FE-AFD-couplings})
is not limited to the linear part. As a matter of fact, the nonlinear
coupling can also have important effects on the spectra \cite{Wang2011b}.
Let us consider a structural phase that has a spontaneous polarization
but no long-range AFD order, which is true above the AFD phase transition
temperature. To simplify the investigation of the dynamics of $\mathbf{u}_{i}$
and $\boldsymbol{\omega}_{i}$ due to their non-linear couplings,
we introduce $\tilde{\mathbf{u}}_{i}$ and $\tilde{\boldsymbol{\omega}}_{i}$
such as:

\begin{align}
 & \begin{cases}
\mathbf{u}_{i}\left(t\right)=\left\langle \mathbf{u}\right\rangle +\tilde{\mathbf{u}}_{i}\left(t\right) & ,\\
\boldsymbol{\omega}_{i}\left(t\right)=\left\langle \boldsymbol{\omega_{R}}\right\rangle +\tilde{\boldsymbol{\omega}}_{i}\left(t\right)=\tilde{\boldsymbol{\omega}}_{i}\left(t\right) & .
\end{cases}\label{eq:fluctuations-R3m}
\end{align}
In the above equation, $t$ represents time and $\tilde{\mathbf{u}}_{i}$
(respectively, $\tilde{\boldsymbol{\omega}}_{i}$) is the deviation
of the local mode (respectively, AFD mode) in the unit cell $i$ with
respect to its spontaneous value $\left\langle \mathbf{u}\right\rangle $
(respectively, $\left\langle \boldsymbol{\omega_{R}}\right\rangle $),
and $\left\langle \boldsymbol{\omega_{R}}\right\rangle $ is the $R$
point average {[}see Eq. (\ref{eq:R-average}){]} oxygen octahedra
tilting, which is zero in the temperature range of investigation.
Plugging Eq. (\ref{eq:fluctuations-R3m}) into Eq. (\ref{eq:FE-AFD-couplings}),
it can be shown that the most important coupling term has the following
form:

\begin{align}
H_{\textrm{coupl,dynam}}= & \sum_{i}\kappa\left\langle u\right\rangle \tilde{u}_{i}\left(\tilde{\omega}_{i}\right)^{2},\label{eq:Ec-nonlinear}
\end{align}
and the dynamical equation for $\tilde{u}_{i}$ is: 
\begin{align}
\frac{d^{2}\tilde{u}_{i}}{dt^{2}}= & -4\pi^{2}\left(\nu_{r}^{\textrm{FE}}\right)^{2}\tilde{u}_{i}-\frac{\kappa\left\langle u\right\rangle }{m_{u}}\left(\tilde{\omega}_{i}\right)^{2}+\frac{Z^{\ast}E(t)}{m_{u}},\label{eq:non-linear-coupled-springs}
\end{align}
where $\nu_{r}^{\textrm{FE}}$ is the natural frequency of the local
mode when there is no coupling with AFD and $m_{u}$ is the local
mode effective mass. $E(t)$ is an applied $ac$ electric field. Equation
(\ref{eq:non-linear-coupled-springs}) shows the existence of a coupling
between the square of the AFD displacement and the displacement of
the local mode. One can prove that when $\nu_{r}^{\textrm{FE}}$ is
close to \textit{twice} the AFD resonance frequency, Eq. (\ref{eq:non-linear-coupled-springs})
leads to \textit{two} resonant frequencies given by $\nu_{r}^{2}=\nu_{\textrm{FE}}^{2}\pm\Omega^{2}$,
where $\Omega^{2}$ depends on the $\kappa$ coupling parameter, as
well as, on the value of the spontaneous polarization, $\left\langle u\right\rangle $.

The nonlinear coupling shown in Eq. (\ref{eq:Ec-nonlinear}) induces
novel dynamical phenomenon in Pb(Zr,Ti)O$_{3}$ solid solutions: in
the simulations, two $A_{1}$ modes appear at a temperature that coincides
with the temperature for which the frequency of the single $A_{1}$
mode equals twice of that of the AFD natural frequency. These two
modes persist for lower temperature. This unexpected doubling of the
$A_{1}$ mode was confirmed by Raman scattering techniques \cite{Wang2011b},
and is a signature of the so-called Fermi resonance, which occurs
when the frequency of one dynamical variable is close to the first
overtone of another dynamical variable \cite{Fermi}.

\subsection{Terahertz Dynamics of multiferroic BiFeO$_{3}$}

With the addition of magnetic moments as another dynamical degree
of freedom, multiferroic BiFeO$_{3}$ is a complicated system. Using
the method detailed in Sec. \ref{sub:Treating-magnetic-moments},
we performed first-principles-based MD simulations on this system,
for the spin-canted antiferromagnetic structure (rather than the magnetic
cycloid) \cite{Wang2012}.

The first important problem to address when dealing with the stochastic
LLG equation is to find the damping constant $\lambda$ of Eq. (\ref{eq:stochastic-LLG}).
This can be achieved by comparing the MD results to those of MC. It
was numerically found that, at any temperature, $\lambda$ has little
effect on the spontaneous polarization and oxygen octahedra tilting,
therefore yielding MD results being similar to the MC predictions
for these structural properties for a wide range of damping coefficients.
However, the effect of $\lambda$ can be clearly seen when investigating
\textit{\emph{magnetic}} properties in the multiferroic BFO -- as
consistent with the fact that $\lambda$ ``only'' appears in the
spin equations of motions. It was found that a wide range of $\lambda$
(namely, $1.0\times10^{-4}\leq\lambda\leq1.0\times10^{-1}$) leads
to a satisfactory agreement (i.e., a difference of less than 3\%)
between the MD and MC results at any temperature for the magnetic
structure. A large range of $\lambda$ can therefore be adopted to
obtain equilibrated properties, which makes the MD approach suitable
to model different multiferroic/ferromagnetic bulks or nanostructures
that may have very different damping constants due to different damping
mechanisms \cite{Gilbert2004}.

We then chose $\lambda=1.0\times10^{-4}$ and computed the complex
electric and magnetic susceptibilities of BiFeO$_{3}$ using equations
similar to Eq. (\ref{eq:dielectric-response-from-MD}). A fixed temperature
of 20\,K was selected, for which the crystallographic equilibrium
state is $R3c$ and both the electric and magnetic susceptibilities
were obtained. Four resonance peaks were found in the dielectric susceptibility,
having resonant frequencies of $151\,\textrm{cm}^{-1}$, $176\,\textrm{cm}^{-1}$,
$240\,\textrm{cm}^{-1}$ and $263\,\textrm{cm}^{-1}$. They correspond
to $E$, $A_{1}$, $E$ and $A_{1}$ symmetries, respectively (note
that these are not all the modes appearing in measured Raman or infrared
spectra \cite{Haumont2006,Kamba2007,Fukumura2007,Lobo2007,Rout2009,Lu2010,Palai2010,Porporati2010,Hlinka2011}
due to the limited number of degrees of freedom included in the effective
Hamiltonian). The first two (lowest-in-frequency) peaks were found
to be mostly related to the sole FE degree of freedom incorporated
in the effective Hamiltonian scheme, while the last two peaks have
also a significant contribution from AFD distortions -- as consistent
with Ref. {[}\cite{Hermet2007}{]}. As revealed in Refs. \cite{Wang2011,Weerasinghe2012}
and explained in Sec. \ref{sub:Linear-coupling}, bilinear couplings
between the FE and AFD modes in the $R3c$ phase allow the AFD mode
to acquire some polarity, which explains why these last two peaks
emerge in the dielectric spectra.

Moreover, two peaks can be seen in the \textit{magnetic} susceptibility,
their predicted resonant frequencies being $\sim7\,\textrm{cm}^{-1}$
and $\sim85\,\textrm{cm}^{-1}$, respectively. Since none of the frequencies
coincides with the dielectric resonant frequencies, it was concluded
that they are not electromagnons, but rather ``solely'' magnons.
Moreover, the lowest-in-frequency magnon entirely \textit{\emph{disappears}}
when the spin-canted structure was intentionally removed in BiFeO$_{3}$
by switching off a specific parameter in the simulation. Tracking
the motions of both the ferromagnetic and antiferromagnetic vectors,
the lowest-in-frequency magnon was found to be associated with the
rotation of magnetic dipoles inside the (111) plane (that contains
the polarization), which is consistent with Refs. \cite{Pincus1960,DeSousa2008a}.
The second magnetic peak is associated with fast oscillations of the
magnetic dipoles going \textit{\emph{in-and-out}} of the $\left(111\right)$
plane, as well as, a change in length of the weak FM vector. This
second peak therefore corresponds to the so-called optic antiferromagnetic
mode of Ref. \cite{Livesey2010} and to the high-frequency gapped
mode of Ref. \cite{DeSousa2008a}. Examining the relation between
the second peak and the coupling between magnetic moments and other
structural dynamical variables, the abnormally large frequency of
the second peak was numerically found to originate from \textit{static}
couplings between the $\mathbf{m}_{i}$'s and structural variables,
with these couplings generating a large magnetic anisotropy. It was
further found that AFD distortions can significantly affect the resonant
frequency of this second magnetic peak.

\subsection{Effect of the central mode on dielectric tunability of ferroelectrics
\label{sub:Tunability-of-SrTiO}}

Dielectric tunability is an important quantity to measure how the
dielectric constant of a material responds to an applied direct-current
($dc$) electric field. Large dielectric tunability near room temperature
in GHz/THz range is important for many technological applications,
such as lens antennas, phased array radars, and tunable filters, etc.
To reveal the connection between tunability and internal dynamics,
the effective-Hamiltonian-based MD approach was applied to the study
of SrTiO$_{3}$ (STO) bulk and strained films (under a tensile strain
of $\simeq1.6\%$) \cite{Weerasinghe2013}. In the MD simulations,
$12\times12\times12$ or $14\times14\times14$ supercells were used
for temperatures being in the interval of 900-130\,K, starting from
high temperature and cooling down the systems. In agreement with experiments
\cite{BST-exp-pd2}, there is no phase transition for STO bulk in
this temperature range. For the strained STO film, however, the $Amm2$
orthorhombic phase (with a polarization pointing along the in-plane
{[}110{]} direction) was found to occur below $T_{c}\sim305\,\textrm{K}$,
which is consistent with measurements \cite{Haeni,Nuzhnyy2009}.

The complex dielectric response was first obtained at different direct
current ($dc$), which was then used to compute the tunability, $\tau\left(E,\nu\right)$,
of STO bulk at a given frequency $\nu$, for each biased $dc$ electric
field $\mathbf{E}$, following the formula: 
\begin{eqnarray}
\tau\left(E,\nu\right) & = & \frac{\textrm{Re}\left[\varepsilon_{\alpha\beta}\left(0,\nu\right)\right]}{\textrm{Re}\left[\varepsilon_{\alpha\beta}\left(\mathbf{E},\nu\right)\right]}\label{tunafromsimul-1}
\end{eqnarray}
where $\textrm{Re}\left[\varepsilon_{\alpha\beta}\left(0,\nu\right)\right]$
and $\textrm{Re}\left[\varepsilon_{\alpha\beta}\left(\mathbf{E},\nu\right)\right]$
are the real parts of the complex dielectric functions for no bias
field and under a $dc$ field $E$, respectively, at the frequency
$\nu$.

At $300$\,K, the tunability of STO bulk was obtained, which is relatively
low ($\sim7\%$) and in good agreement with spectroscopic measurements
\cite{exp:STObulk-harden}. The tunability obtained from MD simulations
was then fitted using Landau-Devonshire-theory-based phenomenological
formulas \cite{55-review-tangantsev}. For low fields, the formula
is 
\begin{eqnarray}
\tau\left(E\right) & = & 1+\beta\left(\varepsilon_{0}\varepsilon\left(0\right)\right)^{3}E^{2},\label{lowfieldtuna-1}
\end{eqnarray}
and for high fields the formula is 
\begin{equation}
\tau\left(E\right)=3\left(\varepsilon_{0}\varepsilon\left(0\right)\right)\beta^{\frac{1}{3}}E^{\frac{2}{3}}\label{highfieldtuna-1}
\end{equation}
where $\varepsilon\left(0\right)$ is the zero-field static dielectric
constant, \textbf{$\varepsilon_{0}$ }is the permittivity of free
space, and $\beta$ is a fitting coefficient, which also appears in
the energy expansion in the Landau-Devonshire theory. For STO bulk,
the low-field value of $\beta$ (to be denoted by $\beta_{L}$) was
found to be nearly equal to its high-field value (to be referred to
as $\beta_{H}$), and both of them are close to experimentally extracted
values of Ref. \cite{55-review-tangantsev,kuzel,exp:STObulk-harden}.

Similar MD simulations were performed at 320K for the epitaxially
strained STO films under a $dc$ field applied along the pseudo-cubic
{[}110{]} direction to compare with experiments \cite{Haeni,Kadlec2009a}.
It is found that there are \textit{\emph{two}}\emph{ }peaks in the
\textit{\emph{in-plane}} $\varepsilon_{xx}$ dielectric response when
\textit{\emph{no}}\emph{ }bias field is applied, which are the CM
and the SM, as consistent with time-domain THz spectroscopy measurements
\cite{Kadlec2009a,Nuzhnyy2009}. The STO film was found to have a
very large tunability near room temperature \cite{55-review-tangantsev,56-exp:tunaSTO/dso-kuzel,Kadlec2009a}.
What is surprising is that the low fields ($E<4$\,MV/m) value $\beta_{L}$
is two orders of magnitude smaller than $\beta_{H}$ in the film.
This significant difference indicates that the Landau-Devonshire theory
cannot adequately predict tunability of epitaxially strained STO films
under low fields. It was verified that such deviation originates from
the existence of the CM in the STO film, which makes the dielectric
response of a system obeyed Eq. (\ref{eq:5}) rather than Eq. (\ref{eq:DHO}).
Similar results were also found in the dielectric tunability of disordered
(Ba$_{0.5}$Sr$_{0.5}$)TiO$_{3}$ solid solution (that also possesses
a CM in its paraelectric phase \cite{Lisenkov}) at 270\,K and at
$\nu$=10\,GHz. The results of STO film and (Ba$_{0.5}$Sr$_{0.5}$)TiO$_{3}$
bulk therefore show that, in the presence of a CM, the Landau-Devonshire
theory cannot be reasonably applied for fitting tunabilities. A new
equation is necessary for fitting the tunability of STO epitaxial
films and other systems exhibiting CM in general. Using Eq. (\ref{eq:5}),
it is shown analytically that, when CM exists, a quartic term in electric
field is necessary to qualitatively represent the influence of the
CM mode on the tunability. Therefore, the tunability is given by a
polynomial of order 4: 
\begin{eqnarray}
\tau_{2}\left(E,0\right) & \simeq & 1+d_{1}E+d_{2}E^{2}+d_{3}E^{4},\label{eq:new-fitting-1}
\end{eqnarray}
where the $E^{3}$ term was numerically found to be negligible.

\section{Conclusion \label{sec:Conclusion}}

In this work, we have demonstrated that the effective-Hamiltonian-based
molecular dynamics simulation approach is a useful tool for investigating
dynamical properties of perovskites. Via examples of BaTiO$_{3}$,
Pb(Zr,Ti)O$_{3}$, BiFeO$_{3}$ and SrTiO$_{3}$, we have shown that
the dielectric responses and tunabilities are closely connected to
the internal dynamics and inherent microscopic couplings.

Let us finish this review by discussing what possible future research
problems may be tackled using the combination of molecular dynamics
and effective Hamiltonian schemes.

First of all, the investigation of dynamical properties of multiferroics
just started. There are therefore a flurry of activities to be done
in that direction. For instance, it may be interesting to take a close
look at the terahertz electric and magnetic dynamics of BFO when the
magnetic ordering consists of the known cycloid. It may be that this
magnetic ordering does lead to the formation of the so-called electromagnons
since it is supposed to give rise to many magnetic peaks that are
equally spaced in frequency and which can then overlap with dielectric
peaks \cite{Sushkov2008,Pimenov2008,Pimenov2009}. It would also be
exciting to find the \textit{dynamical} magneto-electric coefficients,
that is the parameters quantifying how the application of an $ac$
magnetic field (respectively, electric field) affects the time-dependent
electric polarization (respectively, magnetization).

Second, another field of study may be to investigate the terahertz
dynamics of the so-called ferroelectric relaxors. Some typical signatures
of a ferroelectric relaxor are that its $ac$ dielectric response
peaks at certain temperature, with the position of the peak changing
with frequency, while the system remains paraelectric and cubic down
to the lowest temperature. Note that one well known relaxor, namely
the so-called PMN-PT, had been investigated with the MD method, but
based on the bond-valence model \cite{Shin2005,Grinberg2009,Takenaka2013}
rather than effective Hamiltonian. On the other hand, an effective
Hamiltonian for another specific relaxor, namely disordered Ba(Zr$_{0.5}$,Ti$_{0.5}$)O$_{3}$
(BZT) solid solutions, had been shown to generate many of its anomalous
properties \cite{Akbarzadeh2012}. In particular, it was shown that
BZT contains polar nanoregions. Revealing the dynamics of these regions
should lead to a better knowledge of relaxors, and the MD method described
above may be very useful to reach such goal.

Furthermore, first-principles-based simulations recently predicted
the unusual formation of a vortex structure for their electrical dipoles
below a critical temperature in some ferroelectrics \cite{Naumov2004,Sichuga2011,Louis2012},
which was also experimentally observed \cite{Ivry2010,Nelson2011,Jia2011,McQuaid2011}.
Using molecular dynamics simulations coupled with effective Hamiltonians
to reveal the terahertz dynamics of such formation is also a promising
avenue to pursue in a near future.

We thus hope that the use of the effective Hamiltonian within molecular
dynamics simulations will lead to the discoveries of various effects
in a near future, and will continue to enrich our understanding of
complex dynamical phenomena.

\section*{Acknowledgements}

This work is financially supported by NSF DMR-1066158. L.B. also acknowledges
the Department of Energy, Office of Basic Energy Sciences, under contract
ER-46612, ONR Grants N00014-11-1-0384 and N00014-12-1-1034, and ARO
Grant W911NF-12-1-0085 for discussions with scientists sponsored by
these grants. D.W. acknowledges support from the National Natural
Science Foundation of China under Grant No. 51272204.

\end{document}